# Hydrogels enable negative pressure in water for efficient heat utilization and transfer

Yuxi Liu, Zehua Yu, Xiaowei Liu, Peng Cheng, Yifan Zhao, Peihua Yang,* and Kang Liu*

kang.liu@whu.edu.cn (K.L.); peihua.yang@whu.edu.cn (P.Y.)

**Highlights**

Negative pressure up to –1.61 MPa has been achieved and maintained by hydrogel assisted structures

Strong interactions between water and polymers promote negative pressure generation

Negative pressure-driven streaming potential generator outputs a voltage over 1 V

Proof-of-concept of negative pressure heat pipe has been demonstrated

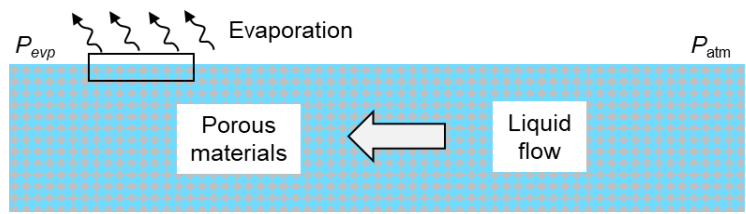

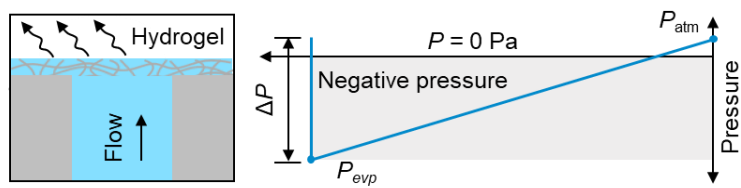

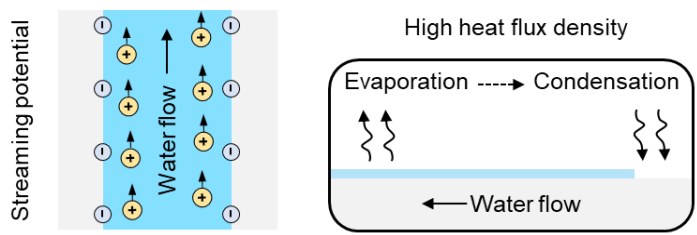

We propose a porous hydrogel assisted structure to generate and maintain absolute negative pressure in water through evaporation at hydrogel surface. Simulations elucidate the essential role of strong interaction between water and polymer network in generating negative pressure. Evaporation-driven electricity generators and heat pipes operating under negative pressure produce sustainable volt-level potential and remarkable heat transfer density, respectively, showing promising advantages and feasibilities in energy utilization and heat transfer.

# Hydrogels enable negative pressure in water for efficient heat utilization and transfer

Yuxi Liu,[1]‡ Zehua Yu,[1]‡ Xiaowei Liu,[1] Peng Cheng,[1] Yifan Zhao,[1] Peihua Yang,[1,2]* and Kang Liu[1]*

[1] MOE Key Laboratory of Hydrodynamic Transients, School of Power and Mechanical Engineering, Wuhan University, Wuhan 430072, China

[2] The Institute of Technological Sciences, Wuhan University, Wuhan 430072, China

‡These authors contributed equally to this work.

* Corresponding to: kang.liu@whu.edu.cn; peihua.yang@whu.edu.cn

ABSTRACT

Metastable water in negative pressure can provide giant passive driving pressure up to several megapascals for efficient evaporation-driven flow, however, the practical applications with negative pressure are rare due to the challenges of generating and maintaining large negative pressure. In this work, we report a novel structure with thin hydrogel films as evaporation surfaces and robust porous substrates as the supports, and obtain a high negative pressure of –1.61 MPa through water evaporation. Molecular dynamics simulations elucidate the essential role of strong interaction between water molecules and polymer chains in generating the negative pressure. With such a large negative pressure, we demonstrate a streaming potential generator that spontaneously converts environmental energy into electricity and outputs a voltage of 1.06 V. Moreover, we propose a "negative pressure heat pipe" for the first time, which achieves a high heat transfer density of 11.2 kW cm$^{-2}$ with a flow length of 1 m, showing the potential of negative pressure in efficient heat utilization and transfer.

INTRODUCTION

Evaporation-driven water flow, which widely exists in nature, has been prevailingly employed in industrial applications such as passive cooling,[1–3] microfluidics chips,[4–6] desalination,[7–9] and recently has attracted increasing interests due to its potential in converting ambient thermal energy into electricity.[10–17] Despite of the diverse applications, evaporation-driven water flow is essentially a process of converting thermal energy into mechanical power of fluidic flow. Driving pressure of the liquid flow, which is the difference between the ambient pressure and the liquid pressure at the liquid/vapor interface, is a key parameter that determines the thermal-to-power efficiency and also the maximum affordable evaporation rate. The lower the pressure at the surface is, the higher the driving pressure is. According to classic nucleation theory,[18,19] the pressure of water can decrease to a metastable state with absolute negative value close to –190 MPa, indicating a possible driving pressure up to 190 MPa. In experiments, previously researchers have also reported observations of negative pressure of –20.3 MPa, –140 MPa and –22 MPa in Berthelot tube, glass slit and hydrogel void, respectively.[20–22] However, the practical application of negative pressure is rarely seen, because of the difficulty in generating and maintaining the negative pressure, especially for an open system with continuous water evaporation and flow.

In a dynamic system, capillarity in porous materials is most commonly used to generate negative pressure.[23,24] According to Young-Laplace equation, liquid pressure at the meniscus decreases to negative when the diameter of pores are hundreds of nanometers. Nevertheless, with the small pore size, negative pressure induced high driving pressure will not naturally produce efficient liquid delivery due to the dramatically increased viscous resistance.[22,25] Xiao et al. proposed to use thin nanoporous membrane at the evaporation surface.[26] The viscous resistance is dictated only by the thickness of the membrane and is decoupled from the driving pressure. However, the thin and porous nature greatly limit the mechanical strength and bonding with supports. Hydrogels are formed through the cross-linking of hydrophilic polymer chains within large amounts of water. The dense and long polymer chains impart hydrogels high mechanical strength. The soft nature and functional groups make the hydrogel facile to bond with substrates.[27–29] More importantly, previous researches have experimentally demonstrated that water enriched hydrogel with proper mechanical rigidity offers ideal interface for evaporation induced negative pressure.[22,30] Hydrogels with intrinsic control over water bring decent platforms to gain insights into negative pressure. However, ideal hydrogel design and practical application scenario with negative pressure is still lacking.

Herein, we report a novel structure with thin hydrogel films as evaporation surfaces and robust porous substrates as the support, to generate negative pressure inside porous substrates for efficient evaporation-driven flow in practical applications (**Figure 1a**). We demonstrate good bonding of hydrogel with different porous materials of diverse sizes and observe a maximum negative pressure value of –1.61 MPa in a glass/hydrogel channel. With molecule dynamic (MD) simulation, we elucidate the role of interaction between water molecules and polymer chains in generating the negative pressure. With the high pressure, we demonstrate the ability of negative pressure in utilization of ambient thermal energy, and prove it with a streaming potential generator that converts environmental energy into electricity. At last, we fabricate a heat pipe with working medium performed at negative pressure for the first time, which achieved a high heat transfer density with a flow length of 1 m, showing the promising potential of negative pressure in efficient energy utilization and heat transfer.

RESULTS AND DISCUSSIONS

We chose poly (2-hydroxyethyl methacrylate) (pHEMA) hydrogel in our work the similar as Wheeler and Wang,[22,31] because of the excellent mechanical property and high capacity of lifting power. **Figure 1b** shows the photograph of the hydrogel thin film, which possess a water content of 28% with an average pore size of 5.87 nm in diameter (**Figure S1**). The pHEMA hydrogel can be fabricated as thin as 50 μm with good softness. The tensile breaking stress reaches 1.62 MPa. To facilitate practical applications, we combined the hydrogel thin film with porous substrates through chemical bonding. The hydrogel film provides the driving pressure, and the substrates support the thin film and ensure low flow resistance, forming a composite structure. **Figure 1c** shows the bonding of hydrogel thin films on sand wick, anodized aluminum oxide (AAO), glass channels and poly-dimethyl siloxane (PDMS) channels. The hydrogels bond well with various substrate materials with porosities ranging from 50 nm to 200 μm. The bonding junction kept stable before the hydrogel film cracks during peeling off (**Figure S2**).

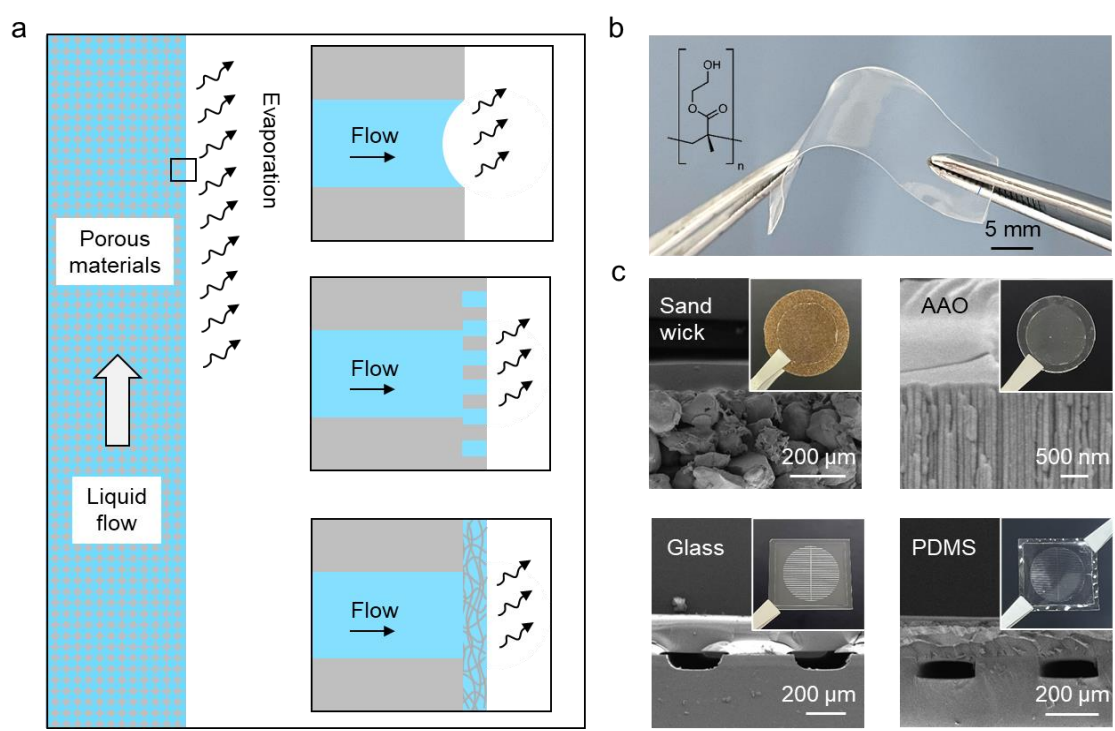

**Figure 1. Structure of the hydrogel porous composites for evaporation-driven water flow.** (a) Schematic of the design with porous medium or thin hydrogel film as evaporation surfaces and porous substrates as the support for evaporation-driven water flow. (b) Typical pHEMA hydrogel thin film with a thickness of 100 μm. Inset is the polymeric structure of pHEMA hydrogel. (c) Cross-sectional view of four hydrogel coated structures. Insets show the original samples.

To evaluate the performance of the composite, we firstly measured the static water lifting power of a hydrogel sand wick (**Figure 2a**). Pore sizes of the sand wick are 25 to 50 μm in diameter (**Figure S3**). The hydrogel sand wick (H-sand wick) could pull the water up to 8.5 m before cavitation (**Figure 2b**). As a comparison, pure sand wick only withstands a height of about 0.75 m, close to the estimated value of 1.14 m calculated from Young-Laplace equation. After we filled the tube with silica powders of 230 μm in diameters, the lifting height increases to 10.04 m. The static lifting pressure comes to 0.1 MPa and the liquid pressure at the water/vapor interface reaches about 0 Pa. Further water lifting induces cavitation occurrence at the bottom surface of the sand wick. Hence, to obtain the maximum lifting power of the hydrogel composites, we designed an osmotic system to take water from saline solution (**Figure S4**). As shown in **Figure 2c**, evaporation from the hydrogel sand wick can continuously take water from saline solution with a concentration of 4.05 mol L$^{-1}$ through reverse osmosis membrane, indicating a driving pressure of 20 MPa. Fitting result reveals that a maximum driving pressure of 22.8 MPa can be achieved.

The hydrogel thin films possess high lifting power up to several tens megapascals, however, cavitation hinders the generation of negative pressure in an evaporation-driven water flow system. According to the analysis from Debenedetti, cavitation is sensitively related to space size, the smaller the size, the more difficult for cavitation to occur.[18] To suppress cavitation, we constructed the hydrogel thin film on glass substrate with fine designed flow channels (**Figure 2d** and **e**). A series of channels with the size of 300 μm × 80 μm serves as the evaporator. A single channel with the size of 32.3 μm × 13.6 μm works as the flow load (**Figure S5**). The length of the load is 15 cm. Under stable evaporation, the driving pressure is calculated by the flow resistance of the channel and evaporation rate. The driving pressure increases with evaporation rate, accompanied by a maximum value of 1.71 MPa (**Figure 2g**). The liquid pressure under the hydrogel film is thus –1.61 MPa, which is the highest value obtained in a flow system. Cavitation happens as the evaporation rate further increases (**Figure 2f**). Compare our results with that of Wheeler's, it is found that, the hydrogel film working as the evaporation surface can generate high negative pressure in water up to several tens megapascals.[22] The difficulty for using the high negative pressure is the cavitation in flow channels. Requirement of the channel property might not be very rigorous. Hydrogel or glass channel are both feasible for implementing negative pressure-driven flow, which suggests loose material constrains for system construction in practical applications. The size of the flow channel is crucial for the negative pressure generation.

Negative pressure-driven flow needs to be confined in micro space to avoid cavitation.

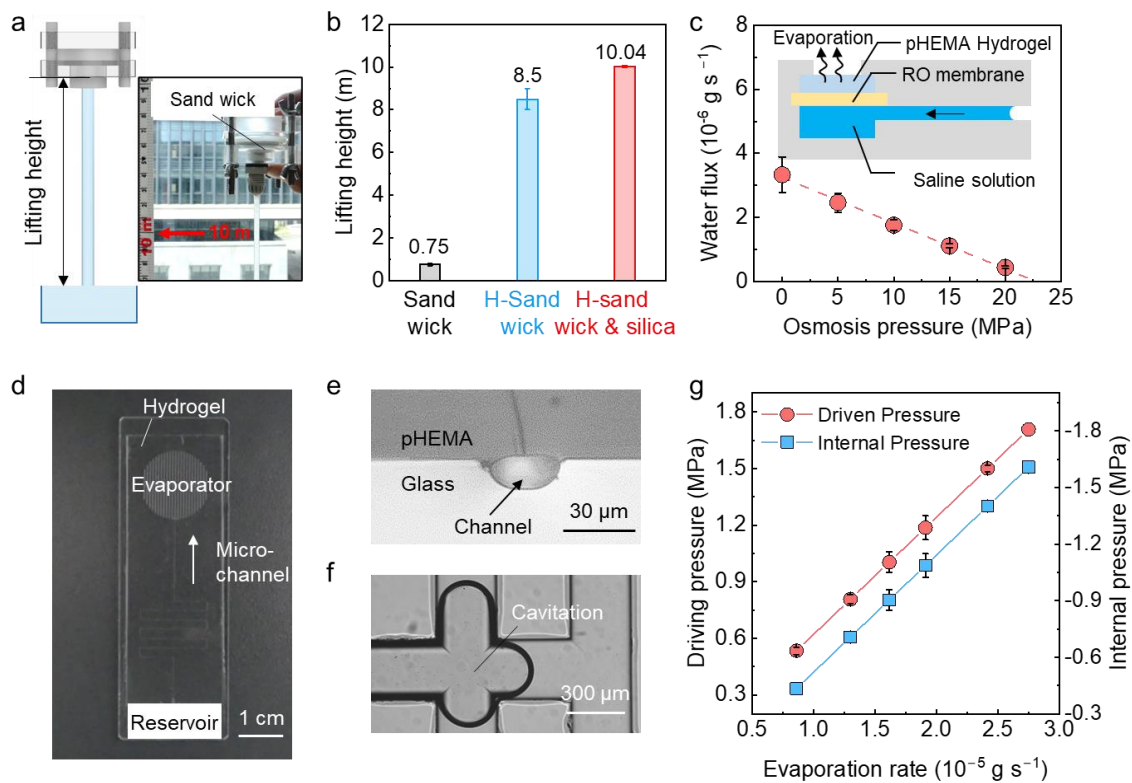

**Figure 2. Negative pressure generated by the hydrogel composites.** (a) Setup of the static water lifting experiment. Inset shows the hydrogel sand wick maintaining water column of 10.04 m without cavitation. (b) Maximum heights of water lifted by sand wick, hydrogel-sand wick with and without silica in the tube. (c) Water adsorption fluxes of hydrogel at different osmosis pressures. Each water flux was collected and averaged over 1-hour test after initial stabilization of 6 hours at 25 °C. Inset shows schematic of the experiment setup. (d) Photograph of the hydrogel-glass microfluidic device. (e) Cross-sectional view of the pHEMA hydrogel bonding on the glass microchannel. (f) Cavitation at the evaporator. (g) Driving pressure and water pressure at the evaporator with different evaporation rates.

To illustrate how the hydrogel generates the negative pressure, we performed MD and density functional theory (DFT) simulations to investigate the water-polymer interaction inside the hydrogel. **Figure 3a** shows the simulated pHEMA hydrogel film and a layer of bulk water. Details about the simulation can been found in **Figure S6, Supporting Note 1 and 2**.[32,33] DFT simulation reveals that, the interaction energy of water-hydroxy group (−5.65 kcal mol$^{-1}$) and water-carbonyl group (−5.23 kcal mol$^{-1}$) are 1.6 and 1.5 times higher than the inter-molecular interaction energy of water (−3.50 kcal mol$^{-1}$), respectively (**Figure 3b**). The interaction scope was characterized by MD simulations through the radial distribution function (RDF) between oxygen in water and that in the functional groups (**Figure 3c**). Results show that, radius of the first hydrated shell of hydroxyl and carbonyl groups in the polymer network is about 0.34 nm. Considering the median pore diameter of 1.19 nm for the pHEMA hydrogel, the volume of the hydrated shell occupies 92.1% of the pore volume assuming a spherical micro pore structure. Hence, most water molecules are strongly interacted with the oxygen-containing functional groups on the polymer chains. According to the current theory of water states in hydrogel,[34–36] we derived that bound and intermediate water take a proportion of 91.1% of total water, also indicating the strong interaction between water and polymer network (Figure 3c inset).

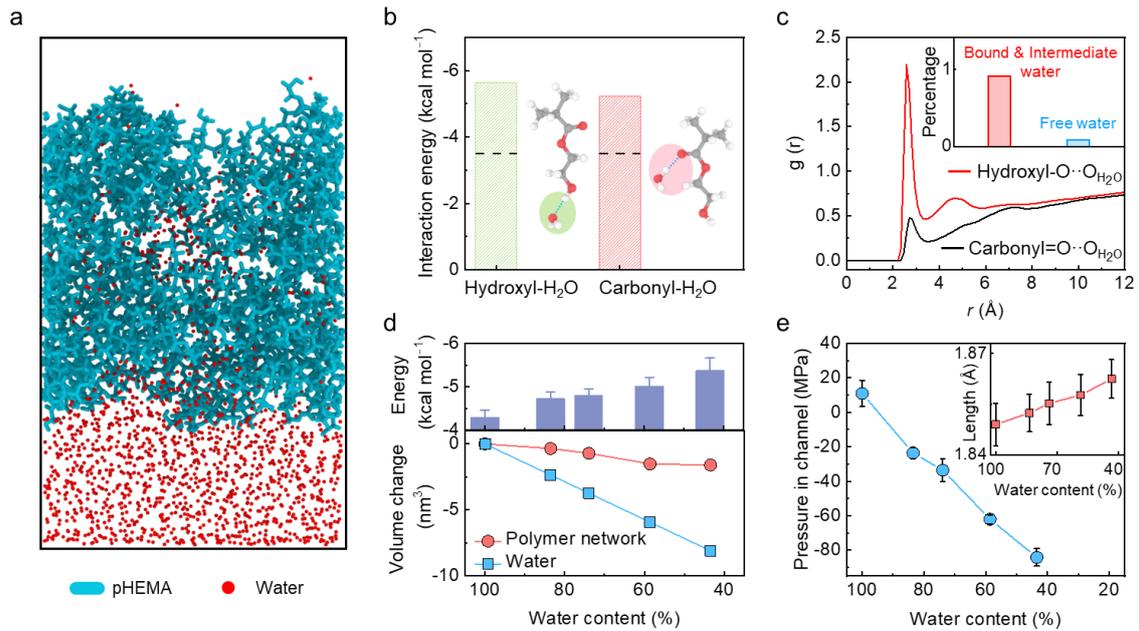

**Figure 3. Mechanism of negative pressure generation.** (a) Snapshot (100% water content) of the hydrogel-water system in MD simulations. (b) DFT calculation of the interaction between oxygen-containing functional groups and water in pHEMA hydrogel. The black dot lines indicate the interaction energy between water molecules. (c) Radial distribution function between oxygen in water and that in hydroxyl groups (red line) and carbonyl groups (blue line). Inset shows the percentage of bound water/intermediate water and free water. (d) Interaction energy between hydrogel network and internal water (top panel), and the volume changes of hydrogel network and evaporated water during evaporation (bottom panel). (e) Negative pressure in bulk water during evaporation. Inset shows the average length of hydrogen bond with water content.

As the water evaporates, the interaction energy between water molecules and hydrogel chains gradually increases. When the water content decreases from 100% to 43%, the interaction energy increases from –4.3 kcal mol$^{-1}$ to –5.4 kcal mol$^{-1}$ (**Figure 3d** top panel, **Figure S7**). Meanwhile, the volume shrinkage of polymer network is much smaller than volume reduction of the evaporated water (**Figure 3d** bottom panel) due to the rigidity of the polymer network. The remaining water molecules undergo a gradually stretching, which induces the migration of outer bulk water into the hydrogel (**Figure S8**). The remaining outer bulk water comes to be stretched subsequently, generating negative liquid pressure in bulk water (**Figure 3e).** As water content in the hydrogel decreases from 100% to 43%, the pressure of bulk water decreases from atmosphere pressure to ca. –84 MPa. The average length of hydrogen bonds confirms the stretching state of bulk water (Figure 3e inset). It is thus revealed that, the strong water-polymer interaction and mechanical rigidity impart the pHEMA hydrogel ability to generate high negative pressure in water.

Negative pressure offers possibility to boost the driving pressure in evaporation-driven water flow, thus can generate high mechanical power output and energy efficiency. To demonstrate the superiority of negative pressure, we fabricated an evaporation-driven streaming potential generator that converts ambient thermal energy into electricity (**Figure 4a**). A piece of hydrogel film covers a glass substrate with embed channels (**Figure 4b**). Evaporation from the top water reservoir spontaneously converts ambient thermal energy into the mechanical energy of water flow in a single microchannel (**Figure S9**). Streaming potential generates due to electric double layer formed at the surface of the glass channel. The output voltage and current were measured through two Ag/AgCl electrodes at the import and outlet of the channel.

As shown in **Figure 4c**, the generator output a voltage of 1.06 V under ambient temperature of 25 ºC and humidity of 40% RH. The voltage decreased dramatically to initial when the device is sealed by a polyethene film, and recovered to the stable value when the plastic film is removed. Such phenomenon can be well repeated. IV curve confirms the electric power generation, and indicates an output current of 0.15 nA (**Figure 4d**). We further varied the ambient humidity and derived the relation between output voltage and driving pressure (**Figure 4e**). The slopes represent surface charge density of –3 mV and –6 mV, which are reasonable for silane modified glass surfaces.[44] The driving pressure at the voltage of 1.06 V is 0.6 MPa, indicating a minimum liquid pressure of –0.5 MPa (in consideration of 0.1 MPa atmospheric pressure). Because of the negative pressure, the evaporation induced voltage with mere streaming potential is comparable to the best values generated by carbon nanomaterials (**Figure 4f**).[12–17] The current density of 29.8 μA cm$^{-2}$ is also among the state-of-the-art values. Connecting three such generators in-series (**Figure 4g** and **Figure S10**), the generators can charge a dielectric capacitor (10 nF) to 2.93 V and power a commercial LED light (**Figure 4h**). It is predicted that, further improvements in the output power can be obtained by increasing the negative pressure, and will achieve a theoretical value of 15 W m$^{-2}$ by utilizing only the ambient heat (**Figure 4i**).

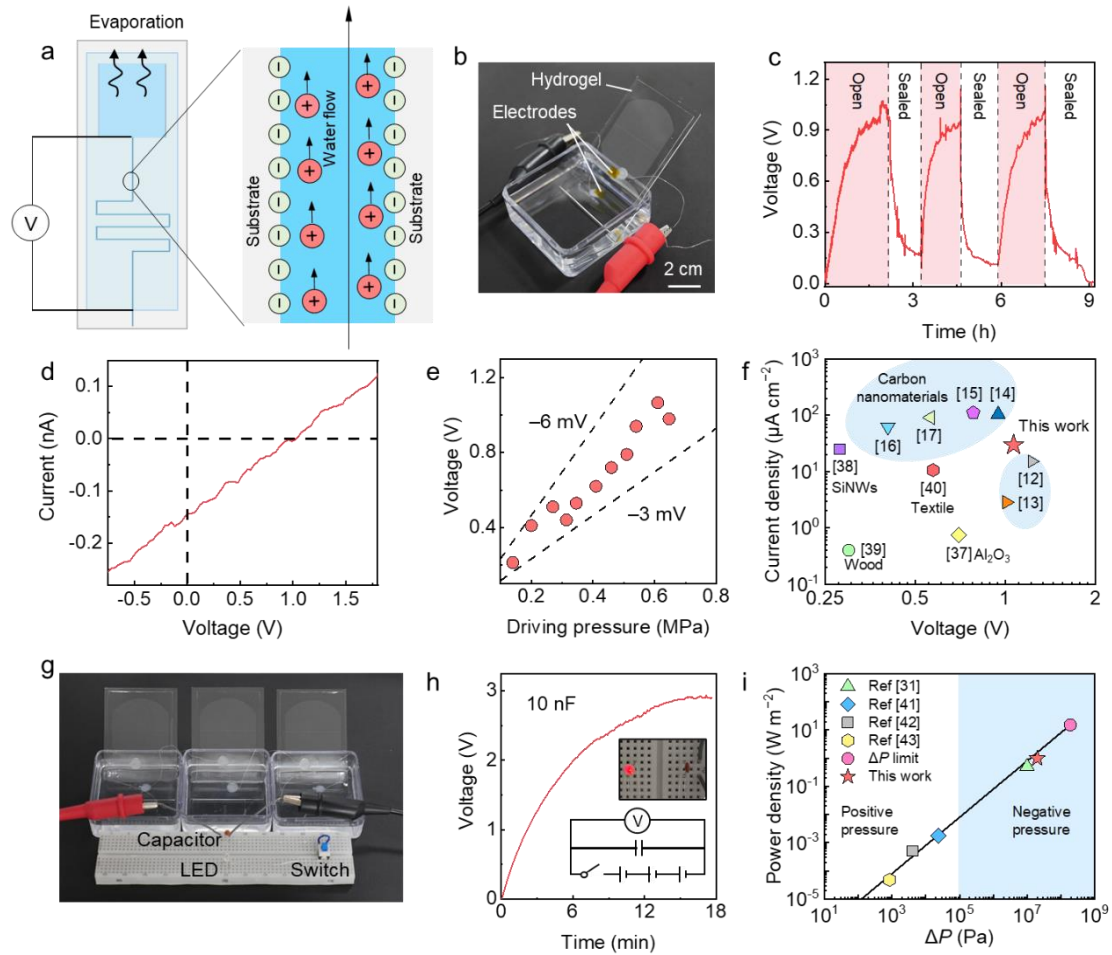

**Figure 4. Evaporation-driven streaming potential generator that converts ambient**

**thermal energy into electricity.** (a) Schematic of the structure and mechanism of the streaming potential generator. (b) Photograph of the device with an evaporation size of 4 cm × 4 cm. (c) Output voltage when the generator was periodically sealed and unsealed. (d) Current-voltage (IV) curve of the generator. (e) Dependence of voltage on driving pressure. Two theoretical dash lines are calculated from $U = \varepsilon_0\varepsilon\zeta\Delta P/(\eta K)$, where $\varepsilon_0\varepsilon$ is the electrical permittivity of water, $\zeta$ is zeta potential, $\eta$ is viscosity, and $K$ is conductivity of water (taken as 2 mS m$^{-1}$). (f) Voltage and current density of the generator comparing with other evaporation-driven generators.[12–17,37–40] (g) Photograph of three generators connected in-series to power an LED light. (h) Voltage-time curve of a 10 nF capacitor charged by the generators. Insets show the equivalent circuit and lighted LED. (i) Mechanical output power via evaporation-driven water flow with different pressure drops.[31,41–43] The black line is calculated from $\Delta P$ times water flux with a constant value of $8 \times 10^{-5}$ kg m$^{-2}$ s$^{-1}$ (natural evaporation at 25 °C and 40% RH).

With the efficient evaporation-driven flow, we also demonstrated its potential in a heat pipe. We fabricated a simplified and visual heat pipe with water (**Figure 5a** and **Figure S11**). Theoretically in a traditional heat pipe, working medium absorbs heat and vaporizes at hot end (route 4′-5′-1′), then diffuses to cold end and releases the absorbed heat while condenses (route 1′-2′-3′) (**Figure 5b**). The condensed water is pumped back to the hot end (3′-4′) by a porous wick. The driving pressure provided by the wick overcome the flow resistance ($\Delta P_c$) to maintain the liquid flow. High heat transfer capacity and long transfer distance will greatly increase the flow resistance and lead to dry out, is thus challenging for heat pipe design. Therefore, we propose to manipulate the liquid under negative pressure (route 3-4-5) to generate higher driving power and overcome larger flow resistance ($\Delta P_n$).

Details of the heat pipe fabrication can be found in **Supporting Method** and **Figure S12**. For easy observation and negative pressure generation, the adiabatic region is only single channel with the hydraulic diameter of 41 μm. The length of the channel is set as 1 m, indicating a heat pipe length of 1 m (**Figure 5c**). As shown in **Figure 5d-f** and **Supporting video 1**, the heat pipe kept stable and no cavitation happened during 4-hour working. We also measured the heat flux transfer capacity of the heat pipe (**Figure 5g** and **h**). Under the temperature difference of 18 °C, the heat flux achieves 11.2 kW cm$^{-2}$ with the single microchannel (**Figure S13, Supporting Note 3**). The liquid pressure under the evaporation surface reaches –1.08 MPa. We termed this device as "negative pressure heat pipe", and the results shows potential of negative pressure heat pipe in high thermal density and long-distance thermal management.

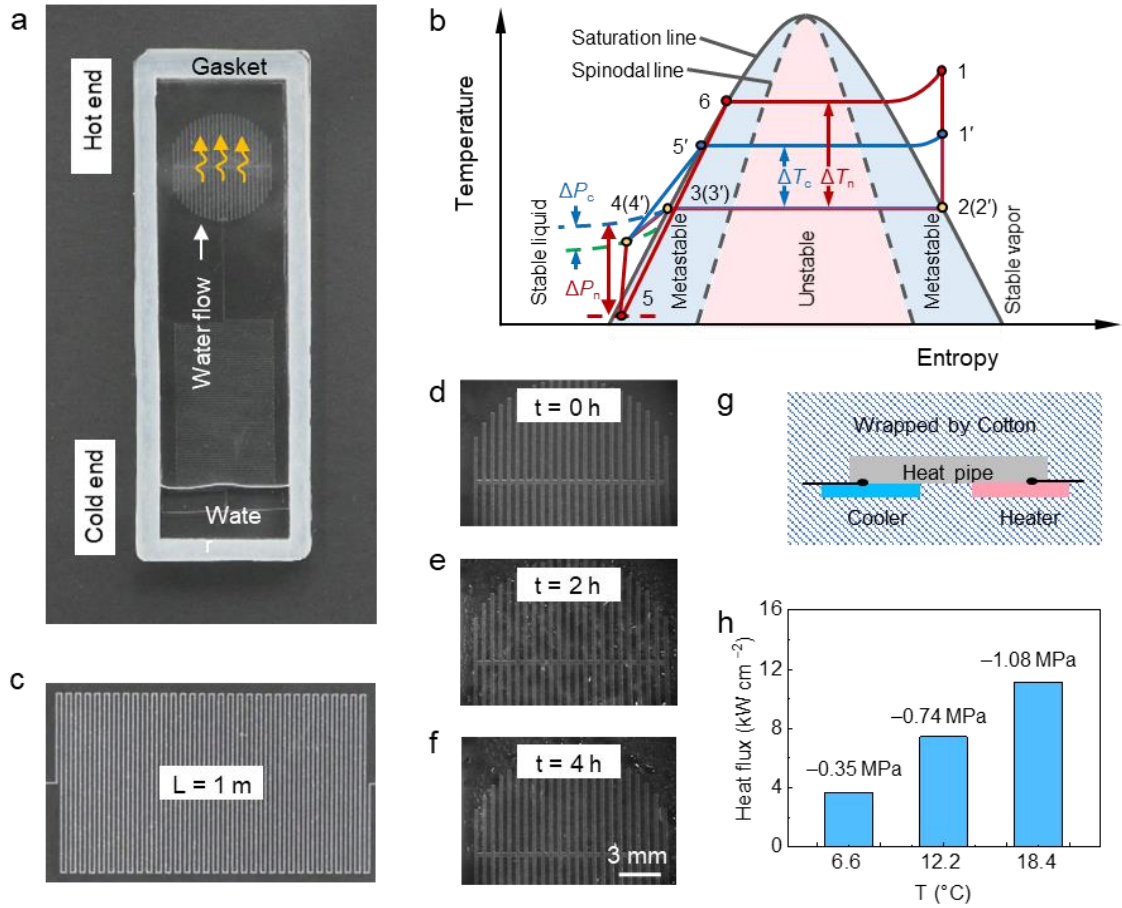

**Figure 5. Negative pressure-driven heat pipe.** (a) Photograph of the simplified and visual heat pipe with the size of 7.5 cm × 2.5 cm × 3 mm. (b) Thermodynamic processes of traditional heat pipe (blue solid line) and negative pressure-driven heat pipe (red solid line). The yellow dots indicate the states that water experiences both in traditional heat pipes and visual heat pipes. The grey solid line and dashed line are the saturation line and spinodal line of water, respectively. The blue and green dashed lines are isobaric lines. (c) Photograph of the enlarged microchannel. (d-f) Snapshots of the evaporation section of the heat pipe during different working time. (g) Schematic of the heat transfer test setup. (h) Heat fluxes transferred by the heat pipe at different temperature differences. Liquid pressure under the evaporation surface in the channel are also given.

CONCLUSIONS

We demonstrate a novel structure composed of thin hydrogel film and porous substrates to generate giant negative pressure in liquid water for efficient heat utilization and transfer. In the composite structure, pHEMA hydrogel was chemically bonded to various porous materials with different sizes and morphologies. Benefiting from strong water-polymer interaction in the pHEMA hydrogel, we observed an evaporation-induced absolute negative pressure of –1.61 MPa in a glass/hydrogel channel. By using the negative pressure, we fabricated a streaming potential generator that spontaneously converts environmental heat energy into electricity and outputs a voltage of 1.06. Moreover, we proposed and demonstrated "negative pressure heat pipe" for the first time, achieving a high heat transfer density of 11.2 kW cm$^{-2}$ with a flow length of 1 m. Therefore, this work may shed insights on the hydrogel enabled negative pressure for practical heat utilization and management, and also gives a promising method to manipulate metastable water in microfluidics.

EXPERIMENTAL PROCEDURES

Resource Availability

*Lead Contact*

Requests for additional information, and for resources and materials should be directed to, and will be fulfilled by, the Lead Contact, Kang Liu (kang.liu@whu.edu.cn).

*Materials Availability*

This study did not generate new unique materials.

Full experimental procedures are provided in the Supplemental Information.

SUPPORTING INFORMATION

Supporting Information includes additional methods, 3 supporting notes, 13 figures and 1 video, and can be found online at xxx.

ACKNOWLEDGMENT

K.L. acknowledges the National Natural Science Foundation of China (51976141, 62161160311). The authors appreciate the Supercomputing Center of Wuhan University for MD/DFT calculations supporting.

AUTHOR CONTRIBUTIONS

K.L. conceived the idea and supervised the project. K.L., P.Y., Y.L. and Z.Y. analyzed the experimental data, drew the figures, and prepared the manuscript. Y.L., P.C., Z.Y., Y.Z., carried out the measurement. Z.Y. performed molecular simulation. Y.L. and Z.Y. provided input for the manuscript. ‡These authors contributed equally to this work.

DECLARATION OF INTERESTS

The authors declare no competing financial interest.

REFERENCES

1. Fang, Y., Chen, G., Bick, M., and Chen, J. (2021). Smart textiles for personalized thermoregulation. Chem. Soc. Rev. *50*, 9357-9374. 10.1039/d1cs00003a.

2. Lu, Z., Strobach, E., Chen, N., Ferralis, N., and Grossman, J.C. (2020). Passive sub-ambient cooling from a transparent evaporation-insulation bilayer. Joule *4*, 2693-2701. 10.1016/j.joule.2020.10.005.

3. Peng, Y., Li, W., Liu, B., Jin, W., Schaadt, J., Tang, J., Zhou, G., Wang, G., Zhou, J., Zhang, C., et al. (2021). Integrated cooling (i-Cool) textile of heat conduction and sweat transportation for personal perspiration management. Nat. Commun. *12*, 6122. 10.1038/s41467-021-26384-8.

4. Bacchin, P., Leng, J., and Salmon, J.B. (2022). Microfluidic evaporation, pervaporation, and osmosis: from passive pumping to solute concentration. Chem. Rev. *122*, 6938-6985. 10.1021/acs.chemrev.1c00459.

5. Seo, S., Ha, D., and Kim, T. (2021). Evaporation-driven transport-control of small molecules along nanoslits. Nat. Commun. *12*, 1336. 10.1038/s41467-021-21584-8.

6. Wong, C.C., Liu, Y., Wang, K.Y., and Rahman, A.R. (2013). Size based sorting and patterning of microbeads by evaporation driven flow in a 3D micro-traps array. Lab Chip *13*, 3663-3667. 10.1039/c3lc50274k.

7. Ghasemi, H., Ni, G., Marconnet, A.M., Loomis, J., Yerci, S., Miljkovic, N., and Chen, G. (2014). Solar steam generation by heat localization. Nat. Commun. *5*, 4449. 10.1038/ncomms5449.

8. Tao, P., Ni, G., Song, C., Shang, W., Wu, J., Zhu, J., Chen, G., and Deng, T. (2018). Solar-driven interfacial evaporation. Nat. Energy *3*, 1031-1041. 10.1038/s41560-018-0260-7.

9. Zhao, F., Zhou, X., Shi, Y., Qian, X., Alexander, M., Zhao, X., Mendez, S., Yang, R., Qu, L., and Yu, G. (2018). Highly efficient solar vapour generation via hierarchically nanostructured gels. Nat. Nanotechnol. *13*, 489-495. 10.1038/s41565-018-0097-z.

10. Cavusoglu, A.H., Chen, X., Gentine, P., and Sahin, O. (2017). Potential for natural evaporation as a reliable renewable energy resource. Nat. Commun. *8*, 617. 10.1038/s41467-017-00581-w.

11. Yin, J., Zhou, J., Fang, S., and Guo, W. (2020). Hydrovoltaic energy on the way. Joule *4*, 1852-1855. 10.1016/j.joule.2020.07.015.

12. Xue, G., Xu, Y., Ding, T., Li, J., Yin, J., Fei, W., Cao, Y., Yu, J., Yuan, L., Gong, L., et al. (2017). Water-evaporation-induced electricity with nanostructured carbon materials. Nat. Nanotechnol. *12*, 317-321. 10.1038/nnano.2016.300.

13. Fang, S., Li, J., Xu, Y., Shen, C., and Guo, W. (2022). Evaporating potential. Joule *6*, 690-701. 10.1016/j.joule.2022.02.002.

14. Ding, T., Liu, K., Li, J., Xue, G., Chen, Q., Huang, L., Hu, B., and Zhou, J. (2017). All-printed porous carbon film for electricity generation from evaporation-driven water flow. Adv. Funct. Mater. *27*, 1700551. 10.1002/adfm.201700551.

15. Liu, K., Ding, T., Li, J., Chen, Q., Xue, G., Yang, P., Xu, M., Wang, Z.L., and Zhou, J. (2018). Thermal-electric nanogenerator based on the electrokinetic effect in porous carbon film. Adv. Energy Mater. *8*, 1702481.

10.1002/aenm.201702481.

16. Yun, T.G., Bae, J., Rothschild, A., and Kim, I.D. (2019). Transpiration driven electrokinetic power generator. ACS Nano *13*, 12703-12709. 10.1021/acsnano.9b04375.

17. Bae, J., Yun, T.G., Suh, B.L., Kim, J., and Kim, I.-D. (2020). Self-operating transpiration-driven electrokinetic power generator with an artificial hydrological cycle. Energy Environ. Sci. *13*, 527-534. 10.1039/c9ee02616a.

18. Debenedetti, P.G. (1996). Metastable Liquids: Concepts and Principles (Princeton University Press).

19. Caupin, F., and Herbert, E. (2006). Cavitation in water: a review. Comptes Rendus Physique *7*, 1000-1017. 10.1016/j.crhy.2006.10.015.

20. Henderson, S.J., and Speedy, R.J. (1980). A Berthelot-Bourdon tube method for studying water under tension. J. Phys. E. *13*, 778-782. 10.1088/0022-3735/13/7/019.

21. Zheng, Q., Durben, D.J., Wolf, G.H., and Angell, C.A. (1991). Liquids at large negative pressures: water at the homogeneous nucleation limit. Science *254*, 829-823. 10.1126/science.254.5033.829.

22. Wheeler, T.D., and Stroock, A.D. (2008). The transpiration of water at negative pressures in a synthetic tree. Nature *455*, 208-212. 10.1038/nature07226.

23. Tas, N.R., Mela, P., Kramer, T., Berenschot, J.W., and van den Berg, A. (2003). Capillarity induced negative pressure of water plugs in nanochannels. Nano Lett. *3*, 1537-1540. 10.1021/nl034676e.

24. Eijkel, J.C., and van den Berg, A. (2005). Water in micro- and nanofluidics systems described using the water potential. Lab Chip *5*, 1202-1209. 10.1039/b509819j.

25. Tyree, M.T. (2003). Plant hydraulics: the ascent of water. Nature *423*, 923. 10.1038/423923a.

26. Xiao, R., Maroo, S.C., and Wang, E.N. (2013). Negative pressures in nanoporous membranes for thin film evaporation. Appl. Phys. Lett. *102*, 123103. 10.1063/1.4798243.

27. Li, W., Liu, X., Deng, Z., Chen, Y., Yu, Q., Tang, W., Sun, T.L., Zhang, Y.S., and Yue, K. (2019). Tough bonding, on-demand debonding, and facile rebonding between hydrogels and diverse metal surfaces. Adv. Mater. *31*, 1904732. 10.1002/adma.201904732.

28. Yuk, H., Zhang, T., Lin, S., Parada, G.A., and Zhao, X. (2016). Tough bonding of hydrogels to diverse non-porous surfaces. Nat. Mater. *15*, 190-196. 10.1038/nmat4463.

29. Gao, Y., Chen, J., Han, X., Pan, Y., Wang, P., Wang, T., and Lu, T. (2020). A universal strategy for tough adhesion of wet soft material. Adv. Funct. Mater. *30*, 2003207. 10.1002/adfm.202003207.

30. Xu, S., Liu, X., Yu, Z., and Liu, K. (2022). Non-contact optical characterization of negative pressure in hydrogel voids and microchannels. Front. Optoelectron. *15*, 10. 10.1007/s12200-022-00016-5.

31. Wang, Y., Lee, J., Werber, J.R., and Elimelech, M. (2020). Capillary-driven desalination in a synthetic mangrove. Sci. Adv. *6*, eaax5253. 10.1126/sciadv.aax5253.

32. Lee, S.G., Brunello, G.F., Jang, S.S., Lee, J.H., and Bucknall, D.G. (2009). Effect of monomeric sequence on mechanical properties of P(VP-co-HEMA) hydrogels at low hydration. J. Phys. Chem. B *113*, 6604-6612. 10.1021/jp8058867.

33. Lee, S.G., Brunello, G.F., Jang, S.S., and Bucknall, D.G. (2009). Molecular dynamics simulation study of P (VP-co-HEMA) hydrogels: Effect of water content on equilibrium structures and mechanical properties. Biomaterials *30*, 6130-6141. 10.1016/j.biomaterials.2009.07.035.

34. Jiang, X., Wang, C., and Han, Q. (2017). Molecular dynamic simulation on the state of water in poly(vinyl alcohol) hydrogel. Comput. Theor. Chem. *1102*, 15-21. 10.1016/j.comptc.2016.12.041.

35. Zhou, X., Zhao, F., Guo, Y., Rosenberger, B., and Yu, G. (2019). Architecting highly hydratable polymer networks to tune the water state for solar water purification. Sci. Adv. *5*, eaaw5484. 10.1126/sciadv.aaw5484.

36. Guo, Y., Bae, J., Fang, Z., Li, P., Zhao, F., and Yu, G. (2020). Hydrogels and hydrogel-derived materials for energy and water sustainability. Chem. Rev. *120*, 7642-7707. 10.1021/acs.chemrev.0c00345.

37. Chi, J., Liu, C., Che, L., Li, D., Fan, K., Li, Q., Yang, W., Dong, L., Wang, G., and Wang, Z.L. (2022). Harvesting water-evaporation-induced electricity based on liquid-solid triboelectric nanogenerator. Adv. Sci. *9*, 2201586. 10.1002/advs.202201586.

38. Qin, Y., Wang, Y., Sun, X., Li, Y., Xu, H., Tan, Y., Li, Y., Song, T., and Sun, B. (2020). Constant electricity generation in nanostructured silicon by evaporation-driven water flow. Angew. Chem. Int. Ed. *59*, 10619-10625. 10.1002/anie.202002762.

39. Zhou, X., Zhang, W., Zhang, C., Tan, Y., Guo, J., Sun, Z., and Deng, X. (2020). Harvesting electricity from water evaporation through microchannels of natural wood. ACS Appl. Mater. Interfaces *12*, 11232-11239. 10.1021/acsami.9b23380.

40. Das, S.S., Pedireddi, V.M., Bandopadhyay, A., Saha, P., and Chakraborty, S. (2019). Electrical power generation from wet textile mediated by spontaneous nanoscale evaporation. Nano Lett. *19*, 7191-7200. 10.1021/acs.nanolett.9b02783.

41. Guan, Y.X., Xu, Z.R., Dai, J., and Fang, Z.L. (2006). The use of a micropump based on capillary and evaporation effects in a microfluidic flow injection chemiluminescence system. Talanta *68*, 1384-1389. 10.1016/j.talanta.2005.08.021.

42. Li, C., Liu, K., Liu, H., Yang, B., and Hu, X. (2017). Capillary driven electrokinetic generator for environmental energy harvesting. Mater. Res. Bull. *90*, 81-86. 10.1016/j.materresbull.2017.02.022.

43. Xu, Z.R., Zhong, C.H., Guan, Y.X., Chen, X.W., Wang, J.H., and Fang, Z.L. (2008). A microfluidic flow injection system for DNA assay with fluids driven by an on-chip integrated pump based on capillary and evaporation effects. Lab Chip *8*, 1658-1663. 10.1039/b805774e.

44. Watson, H., Norström, A., Torrkulla, Å., and Rosenholm, J. (2001). Aqueous

amino silane modification of e-glass surfaces. J. Colloid Interface Sci. *238*, 136-146. 10.1006/jcis.2001.7506.

# Supporting Information for

# Hydrogels enable negative pressure in water for efficient heat utilization and transfer

Yuxi Liu,[1]‡ Zehua Yu,[1]‡ Xiaowei Liu,[1] Peng Cheng,[1] Yifan Zhao,[1] Peihua Yang,[1,2]* and Kang Liu[1]*

[1] MOE Key Laboratory of Hydrodynamic Transients, School of Power and Mechanical Engineering, Wuhan University, Wuhan 430072, China

[2] The Institute of Technological Sciences, Wuhan University, Wuhan 430072, China

‡These authors contributed equally to this work.

* Corresponding to: kang.liu@whu.edu.cn; peihua.yang@whu.edu.cn

**This SI includes:**

**Methods**

**Note 1. Molecular dynamics (MD) simulation**

**Note 2. Density functional theory (DFT) calculations**

**Note 3. Water pressure calculation through heat transfer model**

**Figures**

**References**

**Methods**

**Preparation of pHEMA solution:** The hydrogel fabrication is according to previous literature.[1] The pHEMA hydrogel solution consists of 65 vol% 2-hydroxyethyl methacrylate, 6 vol% ethyleneglycol dimethacrylate, 1 vol% methacrylic acid, 28 vol% de-ionized water and 1 vol% photoinitiator. The photoinitiator was prepared by dissolving 2, 2-dimethoxy-2-phenylacetophenone in N-vinyl pyrrolidone at 600 mg mL$^{-1}$.

**Preparation of modification solution:** A modification solution was prepared by mixing 3 mL 3-(trimethoxysilyl) propyl methacrylate (TMSPMA), 22 μL acetic acid and 150 ml deionized water. After stirring 6 h, a clear solution was obtained.

**Fabrication of hydrogel/porous substrates composites:** The target substrates were first treated with 1 to 10 min oxygen plasma (10.5 W) and then immediately submerged into the as-prepared modification solution for at least 2 h. Afterwards, we pre-polymerized pHEMA hydrogel solution and attached the partial polymerized hydrogel to the modified substrates. Finally, pHEMA hydrogel and the target substrates were fully bonded through additional expose to ultraviolet light (385 nm, 4 mW cm$^{-2}$). In this work, the sand wick was purchased from Wuhan Remdepro Engineering Co., Ltd. The AAO was purchased from Hefei Pu-Yuan Nano Technology Co., Ltd. The etched glass sheet was purchased from Wuhan Jianmizhikong Technology Co., Ltd. The PDMS was synthesized by heating a mixture of 90.9 % [w/w] base and 9.1 % [w/w] curing agent of Sylgard 184 (Dow Corning) for 2 h under 80 °C.

**Peeling off test:** The peeling off tests were conducted by a universal mechanical test machine (CMT6350, SANS). To prepare the test samples, half of the pHEMA hydrogel (100 - 200 μm in thickness, 5 - 6 cm in length and 1.5 -1.8 cm in width) was chemically bonded to the surfaces of substrates, while the other part was set as free ends. Free ends of the hydrogel films were clamped by the chucking appliance of the tensile mechane with 500 N load cells. Samples with rigid substrates were fixed at a platform (**Figure S2**), while samples with soft substrates had both the free ends of hydrogel and substrates clamped. The peeling rates were fixed at 4 mm min$^{-1}$.

**Water lifting test:** We packaged the hydrogel-sand wick composites in a self-made acrylic cell. One side of the cell is open, and the hydrogel faces to atmosphere on this side; the other side of the cell was connected to a water reservoir through a long tube (**Figure 2a**). The tube was filled with water and spherical silica powders, which were used to suppress the cavitation (**Figure S3b**). We placed the water reservoir on the ground and gradually lifted the device to different heights and hovered at each height for an enough time. We recorded the height difference between the device and the water reservoir when the hydrostatic pressure no longer held the water column. The control groups were set without either silica powders or hydrogel (**Figure 2b**).

**Driving pressure test in an osmotic system:** To fabricate the test device, we packaged hydrogel-reverse osmosis (RO) in a self-made acrylic cell. One side of the cell is open, and the hydrogel faces to atmosphere on this side; the other side is connected to a saline reservoir, which is further connected to atmosphere through a horizontal capillary. The reservoir and capillary were filled saline solution with different concentrations (**Figure 2c** inset). The water flux of absorbing pure water from the saline solution via RO membrane in pHEMA hydrogel was obtained by measuring the solution level in the capillary varied with time (**Figure S4**).

**Fabrication of negative pressure heat pipe:** We first etched a microfluidic system, containing a "leaf" for evaporation and a 1 m-long microchannel (hydraulic diameter of 41 μm) for water flow (**Figure S11**), on a glass substrate. Then a thin pHEMA hydrogel film was chemically bonded on the glass substrate, after which the glass/hydrogel composite, a gasket and a smooth glass sheet were sealed together (**Figure S12**). Finally, we injected water into the sealed device with a syringe for the test of the heat pipe.

**Characterizations:** Thermogravimetry analysis was performed via a thermogravimetric analyzer (Mettler-Toledo, TGA2) from 40 to 600 °C with a scan rate of 10 °C min$^{-1}$ under nitrogen atmosphere. Cross section morphologies of pHEMA hydrogel and the displayed hydrogel-coated composites in **Figure 1c** were characterized by a scanning electron microscopy (SEM, SIGMA, Zeiss AG) and optical microscope (NIKON Instruments Co., Ltd.). The pore size of pHEMA hydrogels were characterized through a gas sorption analyzer (Anton Paar, Autosorb-iQ). BET testing conditions and data reduction methods are similar to our previous work.[2] The cross section of the flow channel coated by pHEMA hydrogel film in **Figure 2c** was shot by an inverted microscope (Olympus IX73). The exact microchannel sizes of the etched glass sheet and the microgrooves were measured by a 3D Optical Surface Profiler (NewView™ 9000, Zygo Corporation).

**Supporting Note 1. Molecular dynamics (MD) simulation**

All atom molecular dynamics simulation was performed using the MD code LAMMPS. The network structure was constructed through a 2×2×2 periodic unit array. According to the molar ratio of HEMA and EGDMA in synthesis, each unit contains 34 monomers and 2 crosslinkers. The 2 crosslinkers act as crosslinking sites and there are no self-loops in the structure.[3–6] In order to build a thin hydrogel film, we sealed the end of the polymer chains in z axes with hydrogen to avoid the periodicity in this direction. The system was then solvated with enough water (5980) to simulate the swelling environment of hydrogels.

Since the initial structure was constructed with stretched chain conformations, the first equilibrium process was carried out within isothermal-isobaric (NPT) ensemble at one atmospheric pressure and 300 K until the pressure, density, total energy of the whole system no longer changes. Then an annealing procedure introduced by Lee[4] was applied in order for the system to relaxation out of any possible significant local minima. The resulting pore size is about 2 nm, which is consistent with BET results. Following the annealing procedure, there were 2 layers of water on both sides of the hydrogel, we removed water molecules on one side to simulate the interface of evaporation, and added a virtual wall at the bottom of the other side layer of water to mimic a closed microchannel (bulk water). At interface of hydrogel and bulk water, 1-nm-thick pHEMA hydrogel was fixed to mimic the constrained motion of hydrogel by substrate glass in experiments. The virtual wall was fixed as well. Then a subsequent 10 ns NVT equilibrium is performed to ensure thermodynamic properties were stable.

Afterward, water molecules in the hydrogel gradually evaporated out of the system at a rate of 20 water molecules per ns using the method as introduced in our previous work.[2,7,8] For each evaporation of 20 molecules, the system was performed 8 ns in NVT ensemble, and the last 1 ns of which was recorded for data analysis.

All the MD simulations were performed in periodic boundary conditions and the particle-particle particle-mesh (PPPM) solver was used to calculate the long-range electrostatic interactions in LAMMPS. The cutoff radius was set as 12 Å to calculate van der Waals and short-range electrostatics forces. COMPASS force-field[9] was adopted to describe the bonded and non-bonded interactions among the polymer network and water molecules. A Nose-Hoover thermostat was employed to maintain a constant pressure and temperature in the system. All the simulations were performed using 1 fs time step to integrate the equation of motion. The SHAKE algorithm was applied to keep the water molecular rigid. The coordinates of the system were saved at the interval of a 10 ps simulation. Radius of gyration, pressure of water, and RDF were calculated through LAMMPS commands. In statistics of hydrogen bonds, when distance between donor and acceptor is less than 0.35 nm and simultaneously hydrogen-donor-acceptor angle is less than 30°, it is considered that a hydrogen bond is exist.[10–12] Further analysis and post-processing the simulation trajectories were performed using VMD[13] and OVITO.[14]

The following formula gives the definition of the radius of gyration ($R_g$):

$$R_g = \sqrt{\frac{\sum_{i=1}^{N} m_i (r_i - r_c)^2}{\sum_{i=1}^{N} m_i}} \quad (1)$$

where $N$ is the total number of atoms belong to pHEMA hydrogel, $m_i$ and $r_i$ are mass and vector of site $i$ respectively, $r_c$ is the vector represents the center of mass of the polymer network.

**Supporting Note 2. Density functional theory (DFT) calculations**

The structures of HEMA, water and their complexes (HEMA with water, water with water) were first optimized by using the DFT at the ωB97XD/6-311G (d,p) level.[15,16] Geometry optimizations were performed in implicit solvent model (with SMD) through Gaussian 09 package (Gaussian09 D01., Gaussian Inc., Pittsburgh, PA, 2009). Then the single-point energies of 3 complexes were calculated at the wb97xd/jul-cc-pVTZ level after the optimization, which considering basis set superposition error (BSSE).

The interaction energy ($E_{inter}$) was calculated by the following equation:

$$E_{inter} = E_{AB} - (E_A + E_B) + E_{BSSE} \qquad (2)$$

where $E_A$ and $E_B$ respectively represents the energies of molecular A and B, $E_{BSSE}$ is the BSSE corrected energy and the $E_{AB}$ is the total energy of the complex molecular, smaller the negative value of $E_{inter}$ correspond to stronger interaction.

**Supporting Note 3. Water pressure calculation through heat transfer model**

Theoretically, the thermal resistance of a given material is calculated by $R=\delta/\lambda A$, where $R$ is the thermal resistance, $\delta$ is the thickness along the heat flow, $\lambda$ is the thermal conductivity, $A$ is the cross-sectional area perpendicular to the path of heat flow. The rate of heat transfer of heat flow can be calculated through $\Phi=\Delta T/R$, where $\Phi$ is the rate of heat transfer, $\Delta T$ is the temperature difference across the heat flow.

In this work, the thermal resistance model of the fabricated device is given in **Figure S13**. The thermal conductivity of the used glass sheet and joint gasket are taken as 1.34 K W$^{-1}$ and 0.222 K W$^{-1}$, respectively. Thermal resistances of the glass sheet and the joint gasket could be calculated. Thermal resistances of the coated hydrogel and the joint gasket along the device are much larger than the thermal resistances of glass sheets, phase change and the joint gasket perpendicular to the device. Therefore, the heat transferred through the hydrogel and joint gasket along the device can be neglected. As a result, the rate of heat transferred by evaporation at the pHEMA hydrogel surface can be expressed as:

$$\Phi_e = \Phi_t - \Phi_{w-1} - \Phi_{w-2} = \Phi_t - \frac{\Delta T_t}{R_{w-1}} - \frac{\Delta T_t}{R_{w-2}} \qquad (3)$$

where $\Phi_e$, $\Phi_t$, $\Phi_{w-1}$ and $\Phi_{w-2}$ are the rate of heat transferred by evaporation, the whole device, the substrate wall and the sealing wall, respectively. $\Delta T_t$ is the total temperature difference between the hot end and the cold end of the heat pipe. $R_{w-1}$ and $R_{w-2}$ are the thermal resistances of the glass sheets along the device.

The pressure drop along the microchannel could be calculated by Hagen-Poiseuille equation:[1]

$$\Delta P = \frac{128\mu L}{\rho \pi D^4}\dot{m} \qquad (4)$$

where $\Delta P$ is the pressure drop along the load microchannel, $\mu$ is the viscosity of liquid water (0.9042×10$^{-3}$ kg m$^{-1}$ s$^{-1}$), $L$ is the length of the load microchannel, $\rho$ is the density of liquid water (1000 kg m$^{-3}$), $D$ is the hydraulic diameter of the load microchannel, $\dot{m}$ is the mass rate of water flow measured during evaporation. Then, the evaporation flux can be calculated as $\dot{m}=\Phi_e/\Delta H$, where $\Delta H$ is the evaporation enthalpy.

Consequently, the pressure dropping along the heat pipe microchannel can be calculated.

**Supporting Figures**

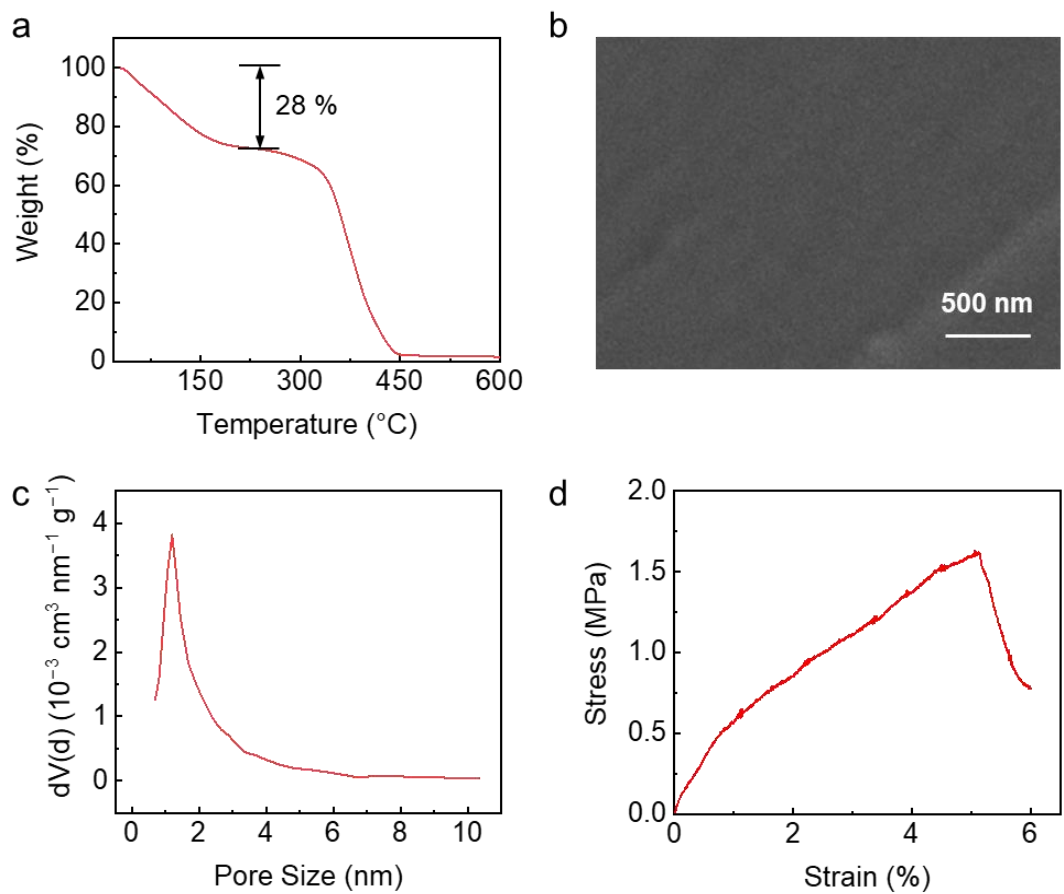

**Figure S1. Characterization of the pHEMA hydrogel.** (a) TGA result. (b) SEM morphology. (c) Pore size distribution in the pHEMA hydrogel. The pore diameters are calculated based on density functional theory (DFT) method. BET result shows that the average and median pore diameter of pHEMA hydrogel are 5.87 and 1.19 nm, respectively. (d) The stress-strain curve of pHEMA hydrogel.

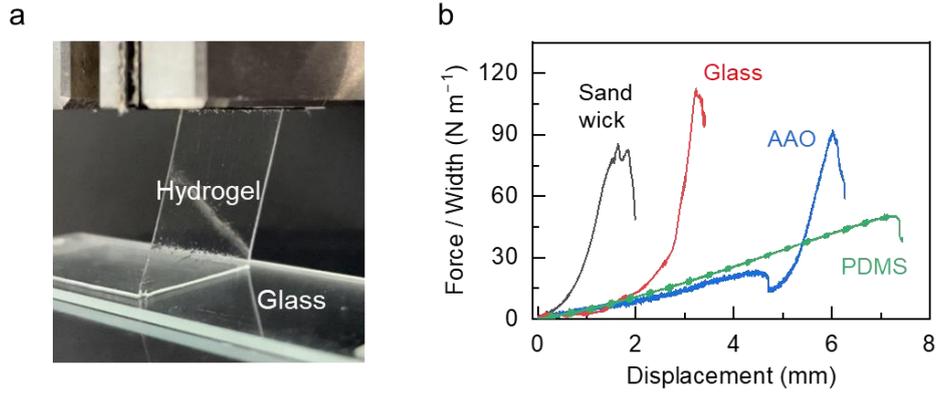

**Figure S2. Peeling off test of hydrogel composites.** (a) The peeling off snapshot of the hydrogel-glass composite. The scale of glass slide is 25 mm × 75 mm. (b) The measured peeling forces per width of pHEMA hydrogel film for various hydrogel-substrates composites.

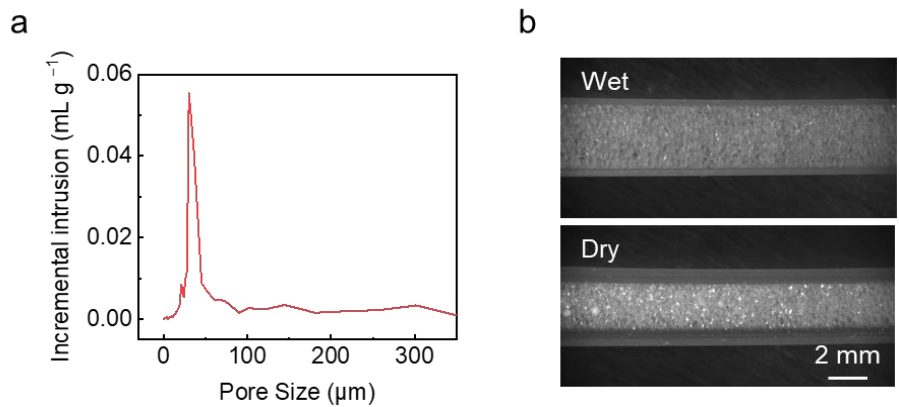

**Figure S3. Hydrogel-sand wick water lifting system.** (a) Pore size distribution in sand wick. Most of the pores distribute between 25 to 50 μm with an average pore diameter of 25.6 μm. (b) Microscope photograph of the silica powder in a tube with and without water.

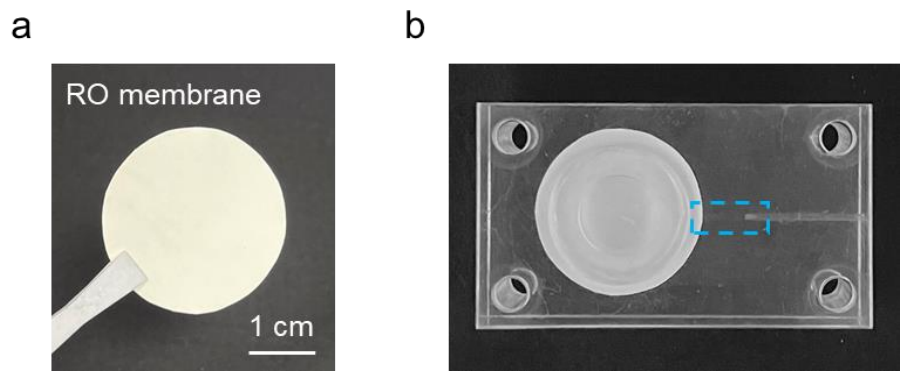

**Figure S4. Hydrogel-RO osmotic system.** (a) Photograph of the RO membrane. (b) Photograph of the hydrogel-RO system with the water column in capillary tube labeled by blue dotted box. Water absorption rate of the system is measured through the water column variation in the transparent 1 mm × 1 mm capillary tube connected to the saline solution reservoir.

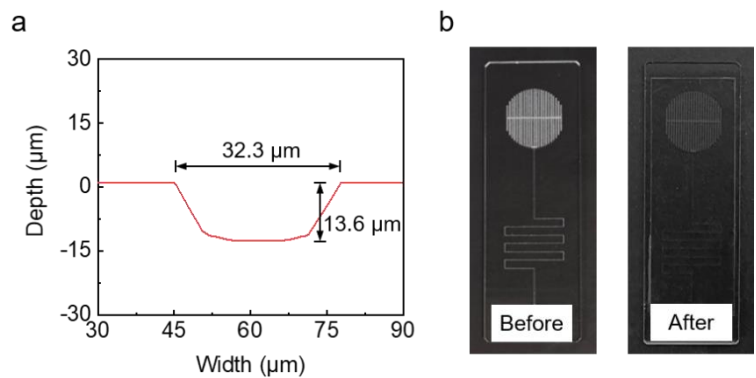

**Figure S5. Hydrogel-glass microfluidic system.** (a) The cross-section profile of the microchannel (32.3 μm × 13.6 μm). (b) The hydrogel-glass microfluidic system before (left) and after (right) covering with hydrogel and filling with water.

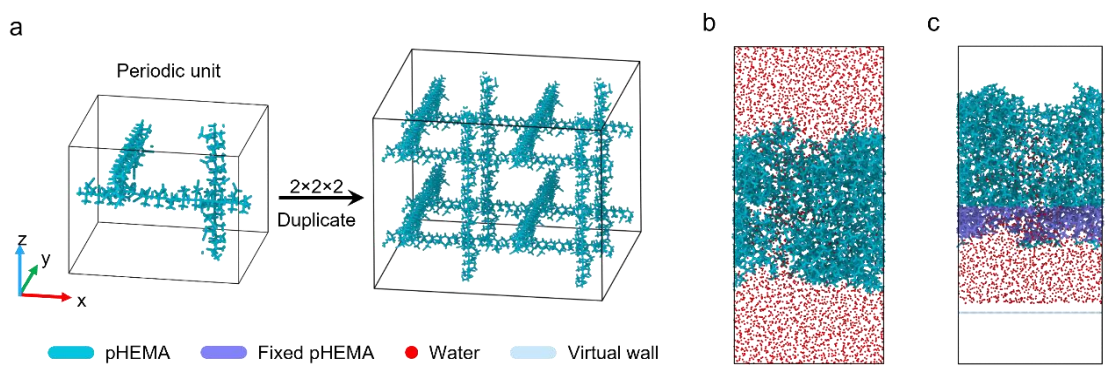

**Figure S6. Construction of the MD model.** (a) Construction of the pHEMA networks. (b) Molecular conformation of pHEMA hydrogel after swelling and subsequent annealing process. (c) Final MD model.

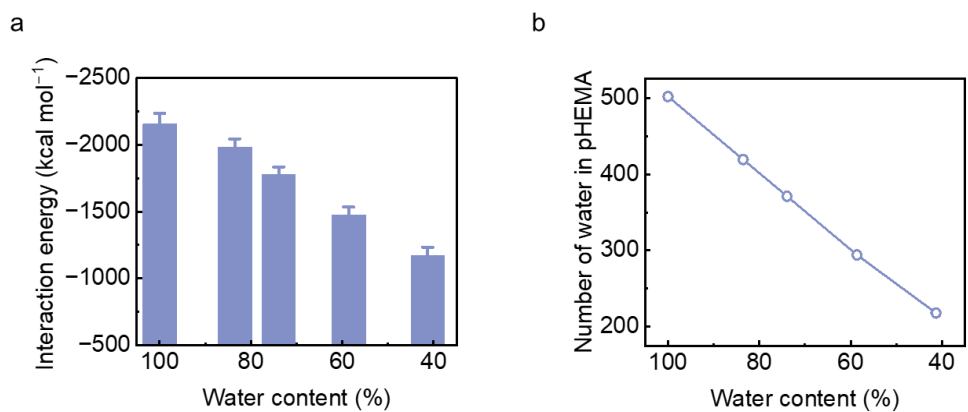

**Figure S7. Interaction energy between hydrogel and water.** (a) Total interaction energy between pHEMA hydrogel and internal water molecules. (b) Total number of water molecules in pHEMA hydrogel network under different water content.

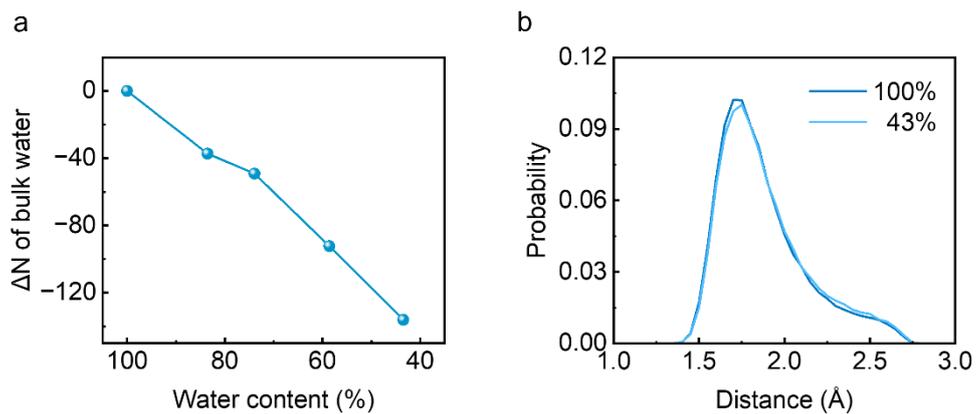

**Figure S8. Stretching of bulk water.** (a) Number of water molecules diffused into the pHEMA hydrogel network from bulk water at different water content of the hydrogel. (b) Hydrogen bond length distribution probability at different water content of the hydrogel.

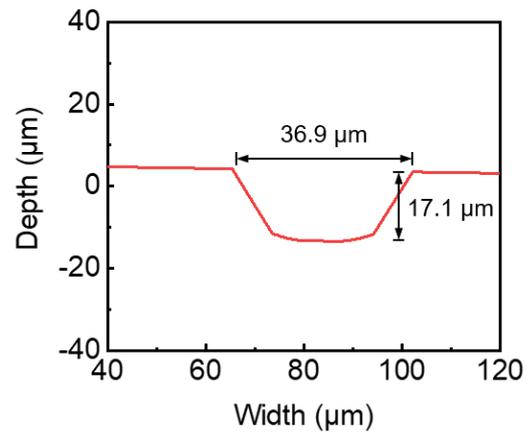

**Figure S9.** The cross-section profile of the microchannel (36.9 μm × 17.1 μm) employed in streaming potential generators, presenting an effective cross-sectional area of 5.03×10$^{-6}$ cm$^2$.

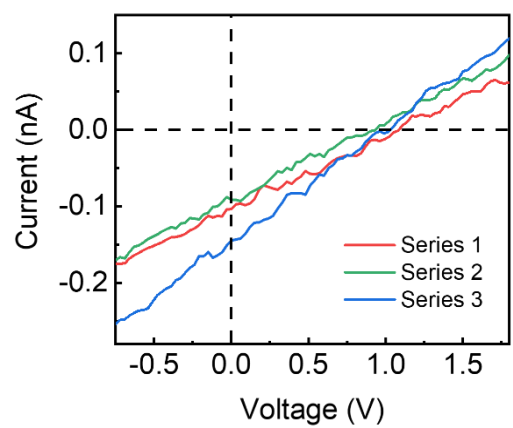

**Figure S10.** Current-voltage curves of three streaming potential generators in series.

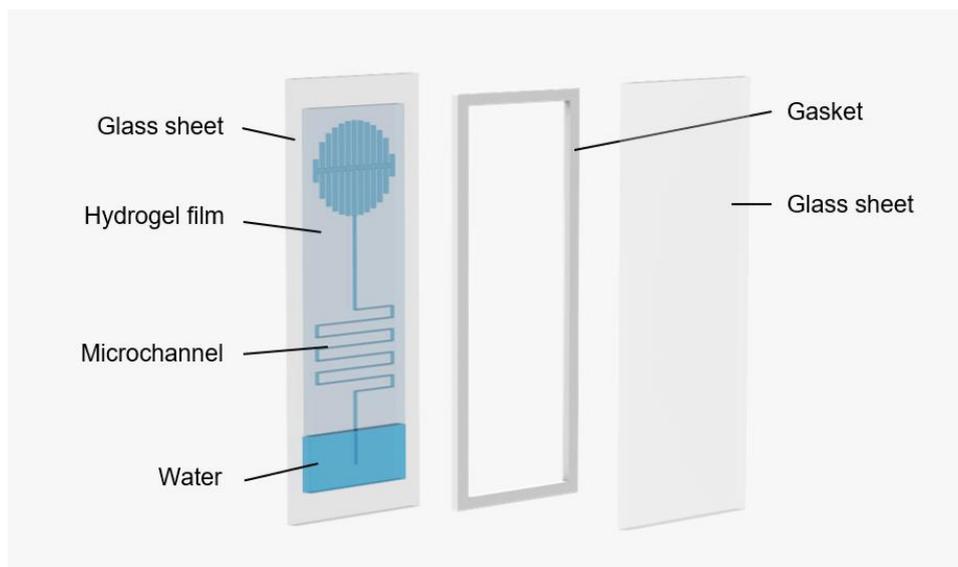

**Figure S11.** 3D explosive view of the fabricated visual heat pipe.

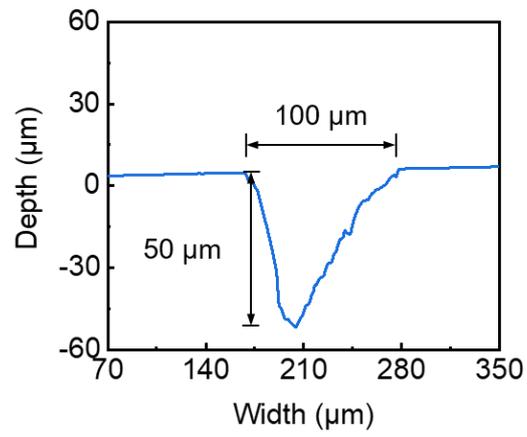

**Figure S12.** The cross-section profile of the microchannel (100 μm width and 50 μm deep) employed in the "negative pressure heat pipe", and an effective hydraulic diameter of 41 μm can be obtained.

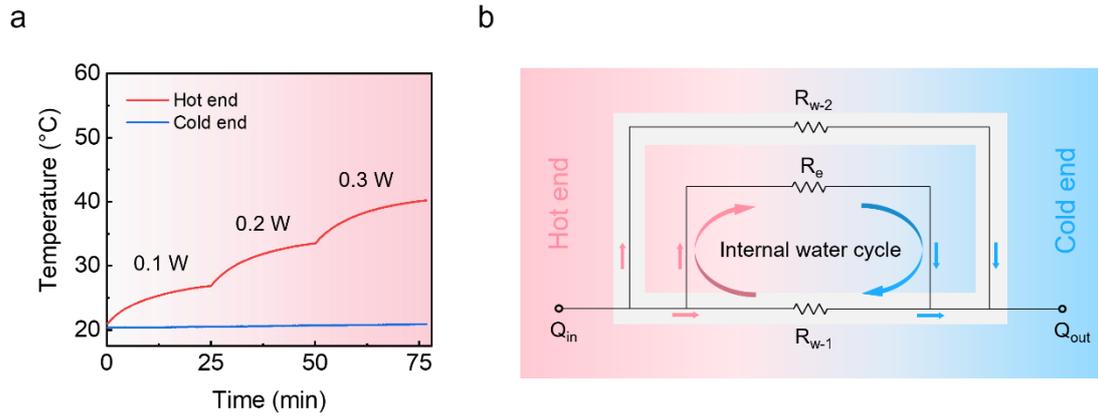

**Figure S13. Test and thermal resistance model of the visual heat pipe.** (a) Temperature of the wrapped heat pipe at different heating powers. (b) Thermal resistance model of the visual heat pipe. There are three heat transfer routes from hot end to cold end of the heat pipe: The first route originates from thermal conduction in tangential direction of glass sheet, joint gasket and hydrogel coating. Thermal resistance of this route is expressed as $R_{w-1}$. The second route originates from thermal conduction in normal direction of joint gasket at hot/cold ends and that in tangential direction of the sealing glass. Thermal resistance of route two is expressed as $R_{w-2}$. The third route originates from the evaporation, diffusion and condensation cycle of the working water. Thermal resistance of route three is expressed as $R_e$. $Q_{in}$ and $Q_{out}$ represents the heat absorbed and released by the heat pipe, respectively.

**References**

1.  Wheeler, T.D., and Stroock, A.D. (2008). The transpiration of water at negative pressures in a synthetic tree. Nature. *455*, 208–212. 10.1038/nature07226.

2.  Liu, Y., Liu, X., Duan, B., Yu, Z., Cheng, T., Yu, L., Liu, L., and Liu, K. (2021). Polymer-water interaction enabled intelligent moisture regulation in hydrogels. J. Phys. Chem. Lett. *12*, 2587–2592. 10.1021/acs.jpclett.1c00034.

3.  Jang, S.S., Goddard, W.A., Yashar, M., and Kalani, S. (2007). Mechanical and transport properties of the poly(ethylene oxide)-poly (acrylic acid) double network hydrogel from molecular dynamic simulations. J. Phys. Chem. B. *111*, 1729–1737. 10.1021/jp0656330.

4.  Lee, S.G., Brunello, G.F., Jang, S.S., and Bucknall, D.G. (2009). Molecular dynamics simulation study of P(VP-co-HEMA) hydrogels: Effect of water content on equilibrium structures and mechanical properties. Biomaterials. *30*, 6130–6141. 10.1016/j.biomaterials.2009.07.035.

5.  Lee, S.G., Brunello, G.F., Jang, S.S., Lee, J.H., and Bucknall, D.G. (2009). Effect of monomeric sequence on mechanical properties of P(VP-co-HEMA) hydrogels at low hydration. J. Phys. Chem. B. *113*, 6604–6612. 10.1021/jp8058867.

6.  Sun, D., and Zhou, J. (2012). Effect of water content on microstructures and oxygen permeation in PSiMA-IPN-PMPC hydrogel: A molecular simulation study. Chem. Eng. Sci. *78*, 236–245. 10.1016/j.ces.2011.11.020.

7.  Liu, X., Wei, W., Wu, M., Liu, K., and Li, S. (2019). Understanding the structure and dynamical properties of stretched water by molecular dynamics simulation. Mol. Phys. *117*, 3852–3859. 10.1080/00268976.2019.1669835.

8.  Wu, M., Wei, W., Liu, X., Liu, K., and Li, S. (2019). Structure and dynamic properties of stretched water in graphene nanochannels by molecular dynamics simulation: Effects of stretching extent. Phys. Chem. Chem. Phys. *21*, 19163–19171. 10.1039/c9cp03981c.

9.  Sun, H. (1998). Compass: An ab initio force-field optimized for condensed-

phase applications - Overview with details on alkane and benzene compounds. J. Phys. Chem. B. *102*, 7338–7364. 10.1021/jp980939v.

10. Luzar, A., and Chandler, D. (1996). Effect of environment on hydrogen bond dynamics in liquid water. Phys. Rev. Lett. *76*, 928–931. 10.1103/PhysRevLett.76.928.

11. Soper, A.K., and Phillips, M.G. (1986). A new determination of the structure of water at 25°C. Chem. Phys. *107*, 47–60. 10.1016/0301-0104(86)85058-3.

12. Teixeira, J., and Bellissent-Funel, M.C. (1990). Dynamics of water studied by neutron scattering. J. Phys. Condens. Matter. *2*, 2–6. 10.1088/0953-8984/2/S/011.

13. Humphrey, W., Dalke, A., and Schulten, K. (1996). VMD: Visual Molecular Dynamics. J. Mol. Graph. *14*, 33–38.

14. Stukowski, A. (2010). Visualization and analysis of atomistic simulation data with OVITO – the Open Visualization Tool. Model. Simul. Mater. Sci. Eng. *18*, 015012. 10.1088/0965-0393/18/1/015012.

15. Kohn, W., and Sham, L.J. (1965). Self-consistent equations including exchange and correlation effects. Phys. Rev. *140*, A1133.

16. Yang, J., Xu, Z., Wang, J., Gai, L., Ji, X., Jiang, H., and Liu, L. (2021). Antifreezing zwitterionic hydrogel electrolyte with high conductivity of 12.6 mS cm−1 at −40 °C through hydrated lithium ion hopping migration. Adv. Funct. Mater. *18*, 2009438. 10.1002/adfm.202009438.